\documentclass[]{interact}

\usepackage{epstopdf}% To incorporate .eps illustrations using PDFLaTeX, etc.
\usepackage[caption=false]{subfig}% Support for small, `sub' figures and tables

\usepackage[numbers,sort&compress,merge]{natbib}% Citation support using natbib.sty
\bibpunct[, ]{[}{]}{,}{n}{,}{,}% Citation support using natbib.sty
\renewcommand\bibfont{\fontsize{10}{12}\selectfont}% Bibliography support using natbib.sty

\usepackage{todonotes} % for the to do notes
\usepackage[version=4]{mhchem} % Chemistry formulas

\theoremstyle{plain}% Theorem-like structures provided by amsthm.sty

\theoremstyle{definition}

\theoremstyle{remark}

\begin{document}

%\articletype{ARTICLE TEMPLATE}% Specify the article type or omit as appropriate

\title{Adsorption from Binary Liquid Solutions into Mesoporous Silica: A Capacitance Isotherm on 5CB Nematogen/Methanol Mixtures}

\author{
\name{Andriy V. Kityk \textsuperscript{a,b}, Gennady Y. Gor\textsuperscript{c} and Patrick Huber\textsuperscript{a,d,e}}
\affil{\textsuperscript{a}Hamburg University of Technology, Center for Integrated Multiscale Materials Systems CIMMS, 21073 Hamburg, Germany;\\ \textsuperscript{b}Faculty of Electrical Engineering, Czestochowa University of Technology, 42-200 Czestochowa, Poland; \\
\textsuperscript{c}Otto H. York Department Chemical and Materials Engineering, 
New Jersey Institute of Technology, University Heights, Newark, NJ 07102, USA;\\
\textsuperscript{d}Deutsches Elektronen-Synchrotron DESY, Center for X-Ray and Nano Science CXNS, 22603 Hamburg, Germany;\textsuperscript{e}Hamburg University, Centre for Hybrid Nanostructures CHyN, 22607 Hamburg, Germany }
}

\maketitle

\begin{abstract}
We present a capacitance method to measure the adsorption of rod-like nematogens (4-cyano-4'-pentylbiphenyl, 5CB) from a binary liquid 5CB/methanol solution into a monolithic mesoporous silica membrane traversed by tubular pores with radii of 5.4\,nm at room temperature. The resulting adsorption isotherm is reminiscent of classical type II isotherms of gas adsorption in mesoporous media. Its analysis by a model for adsorption from binary solutions, as inspired by the Brunauer-Emmett-Teller (BET) approach for gas adsorption on solid surfaces, indicates that the first adsorbed monolayer consists of flat-lying (homogeneously anchored) 5CB molecules at the pore walls. An underestimation of the adsorbed 5CB amount by the adsorption model compared to the measured isotherm for high 5CB concentrations hints towards a capillary filling transition in the mesopores similar to capillary condensation, i.e. film-growth at the pore walls is replaced by filling of the pore centers by the liquid crystal. The experimental method and thermodynamic analysis presented here can easily be adapted to other binary liquid solutions and thus allows a controlled filling of mesoporous materials with non-volatile molecular systems.   
\end{abstract}


\section{Introduction}

Spatial confinement of liquid crystals can alter their physico-chemical properties substantially. Novel phase behaviour, complete suppression of phase transitions and inhomogeneous structures and dynamics have been reported experimentally and validated both by analytical theory and computer simulation\cite{Crawford1996, Binder2008, Kityk2008, Kityk2010, Muter2010,Grigoriadis2011, Araki2011,Zhang2012,Duran2012, Yildiz2015, Huber2015, Ryu2016, Dietrich2017, Klop2018}. In particular the advent of porous media with tailorable pore shapes and tuneable pore size from the macro-, via the meso- to the microscale have resulted in an increased number of studies aimed at an understanding of liquid crystalline behaviour in these interface-dominated geometries \cite{Crawford1996, Lagerwall2012, Schlotthauer2015, Huber2015, Huber2020}.

Moreover, mesoporous media with pores smaller than 50\,nm exhibit structures significantly smaller than the wavelengths of visible light and can thus act as photonic metamaterials. Their optical functionality is not determined by the properties of the base materials, but rather by the precise pore shape, geometry, and orientation. Embedding molecular matter, most prominently liquid crystals in pore space provides additional opportunities to control light-matter interactions at the single-pore, meta-atomic scale \cite{Busch2017, Sentker2018, Yildirim2019,  Sentker2019}. The resulting hybrid materials get their mechanical stability from the porous solid, whereas the liquid-crystalline pore filling adds an integrated functionality to the system. To exploit and predict the potential of liquid crystal-infused solids as functional nanomaterials, however, a detailed understanding of the physico-chemical changes of the confined compared to the unconfined, bulk liquid crystals is necessary \cite{Huber2020}.

Another challenge in the study of liquid crystals in porous media compared to other molecular fillings is their often very low vapour pressure under ambient conditions. Therefore, the liquid crystals are typically filled via capillary action (spontaneous imbibition) into the porous solids \cite{Kityk2008,Gruener2011, Kityk2014}. This preparation scheme precludes a partial filling of pore space, e.g. the adsorption of a thin liquid crystalline film at the pore walls. Hence, almost solely completely with LC filled porous solids have been explored so far \cite{Huber2015, Huber2020}. 

Here we present an experimental study, where we fill sequentially a mesoporous monolithic silica membrane with pores 13\,nm across by imbibition of binary mixtures of methanol and 5CB, 5CB$_{\rm x}$CH$_{\rm 1-x}$ with distinct concentrations $x$ of 5CB. We try to infer the 5CB adsorption after each imbibition step. The resulting filling-fraction versus concentration isotherm is then analysed with a model for adsorption from binary solutions.

\section{Experimental}
A monolithic mesoporous silica SiO$_2$ membrane of 280\,$\mu$m thickness is prepared by thermal oxidation of mesoporous silicon, pSi, at 1073\,K for 12\,h. The pSi membranes are synthesized by electrochemical anodic etching of highly p-doped (100)-oriented silicon wafers employing a mixture of concentrated fluoric acid and ethanol (volume ratio 2:3) as electrolyte \cite{Kumar2008, Sailor2011}. The resulting pSiO$_2$ membranes consist of an array of parallel-aligned nanochannels of mean diameter $D=10.8\pm$0.5\,nm and exhibit a porosity P=35$\pm$2\,\%, as determined by recording a volumetric nitrogen sorption isotherm at 77\,K. The pore diameter is determined from the adsorption branch of the isotherm using the Kelvin method. 

\begin{figure}
    \centering
    \includegraphics[width=0.95\textwidth]{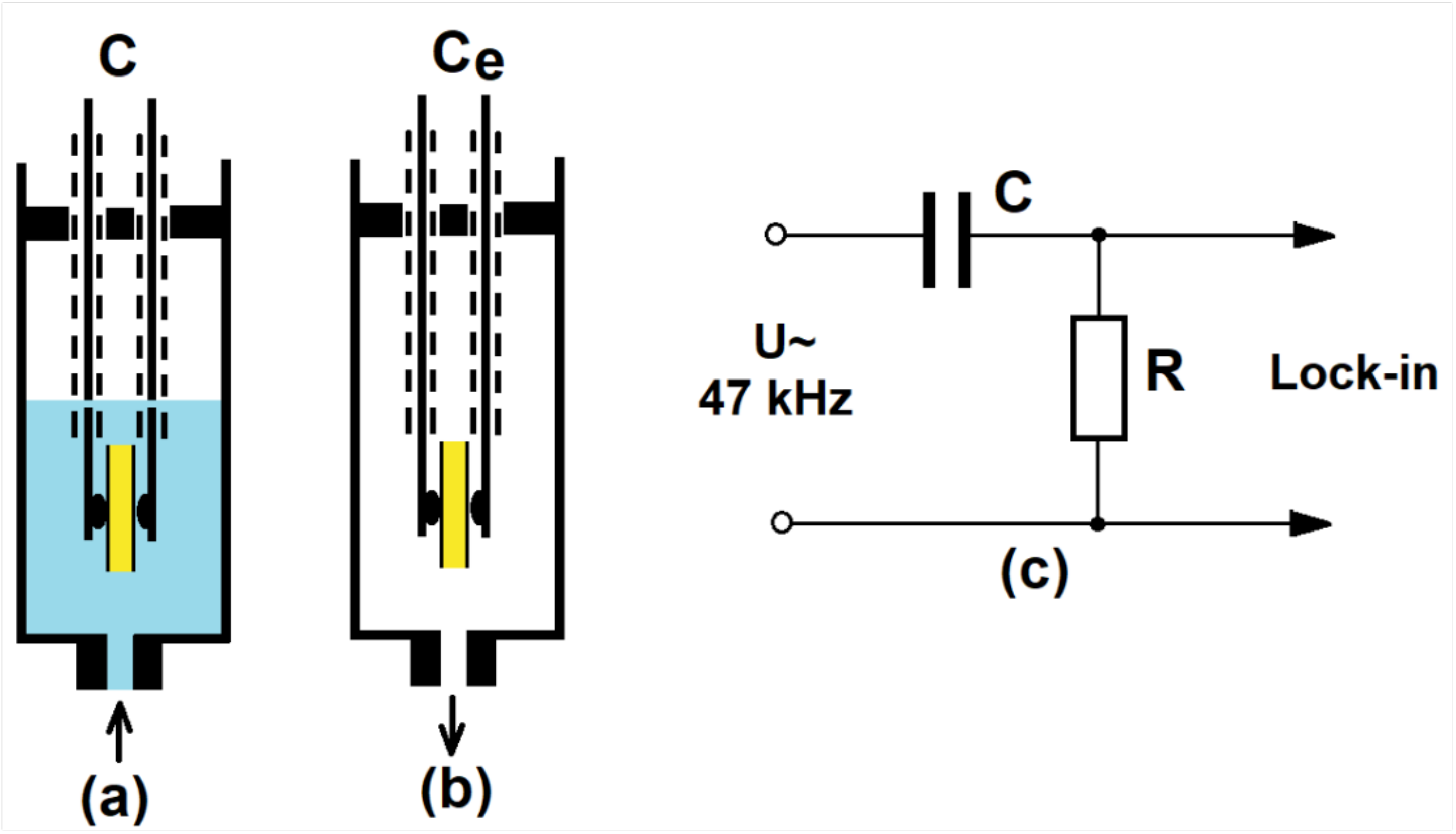}
    \caption{Experimental scheme for the measurement of an electrical capacitance isotherm from binary solutions. A liquid glass container with integrated sample holder for capacitance measurements on a monolithic mesoporous membrane in liquid environment. (a) The adsorption process takes place in the CH($1-x$)5CB($x$) solution filled cell. (b) The capacitance $C_{\rm e}$ is measured immediately after emptying of the liquid container. (c) Electrical circuit used in the capacitance measurements.}
    \label{fig:experimental_setup}
\end{figure}

The experimental details of the electrical capacitance measurements are presented in Fig.~\ref{fig:experimental_setup}. Gold electrodes are deposited on the nanoporous membrane of pSiO$_2$. The capacitance $C$ of the sample is measured by a Lock-in amplifier SR830 in internal reference mode at a frequency of 47\,kHz, see the electric circuit in Fig.~\ref{fig:experimental_setup}(c). In such a setup the measured voltage, $U_\text{R}$ at the resistor $R$, is given by 
\begin{equation}
    U_\text{R}=\frac{RU}{\sqrt{(\omega C)^{-2}+R^2}} \approx \omega RCU
    \label{eq:voltage_U_R}
\end{equation}
The simplification of Eq.~\ref{eq:voltage_U_R} follows from the fact that here $1/(\omega C) \gg R$. The ratio of these values ranges from 20 to 50. Thus the correction to the measured $C$ value due to ignoring of $R$ in the square root denominator is less than 0.001 $C$ at maximum. In this case the capacitance $C \propto U_\text{R}$. Thus one can refer to the measured lock-in $U_\text{R}$ value as capacitance $C$ presented in arbitrary units. The absolute capacitance in our measurement is not important, only the relative change is of relevance. Moreover, the phase shift in the whole range of volume fractions $x$ of  the liquid mixture 5CB$_{\rm x}$CH$_{\rm 1-x}$  embedded into nanoporous matrix changes roughly between 87 and 89\,deg, i.e. by less than 2\,deg only. This means that we are evidently far away from any dielectric relaxation frequencies ($\omega \tau \gg 1$). Otherwise the analysis of the dielectric measurement would be much more complicated, see Ref.~\cite{Calus2015a}.

\section{Results and Discussion}

\begin{figure}
    \centering
    \includegraphics[width=0.85\textwidth]{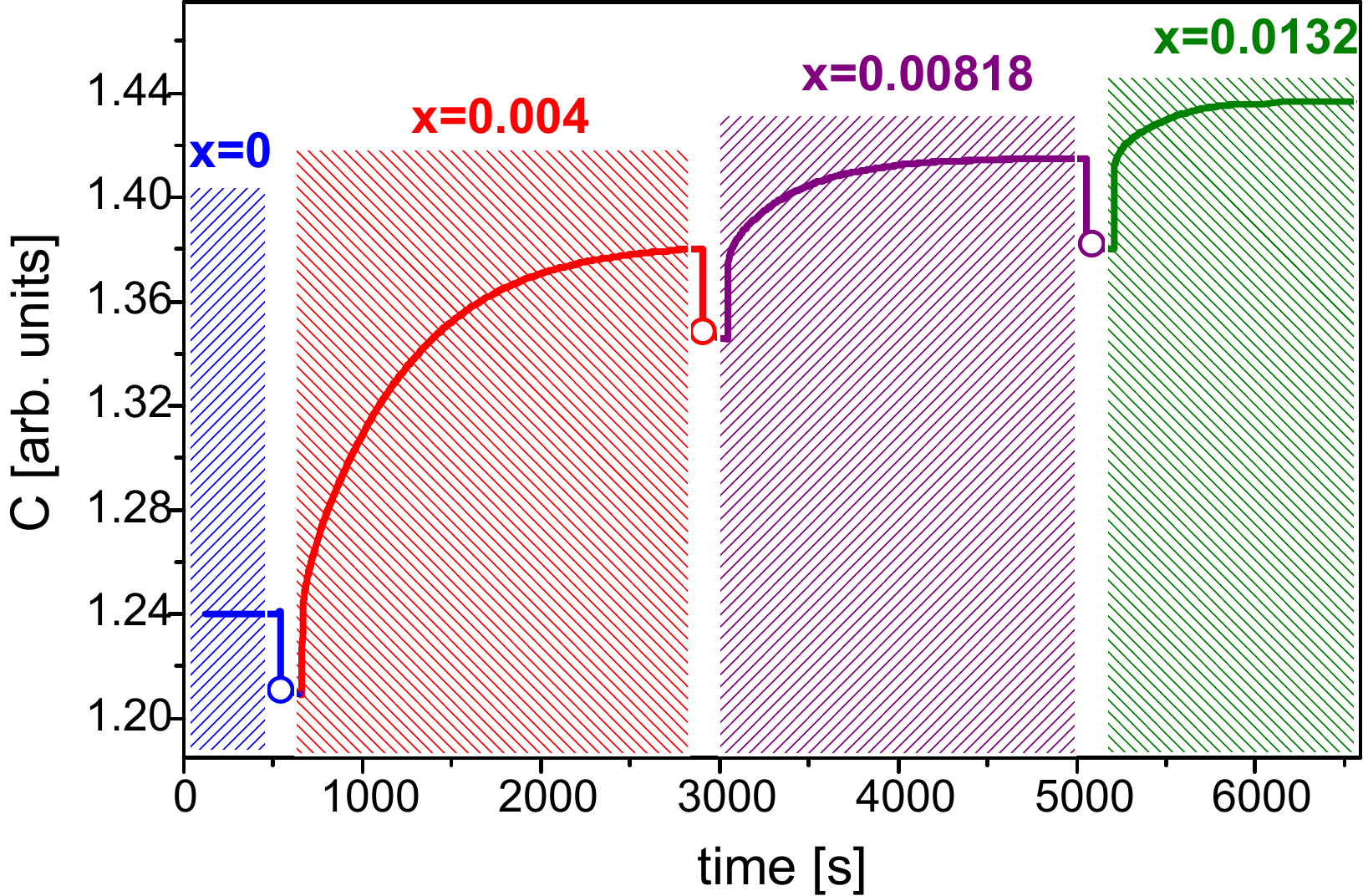}
    \caption{Adsorption of 5CB from binary mixtures 5CB$_{\rm x}$CH$_{\rm 1-x}$ in mesoporous pSiO$_2$ as explored by electrical capacitance measurements. Marked by different colours are temporal intervals during which the 5CB adsorption into the mesoporous host is monitored by \textit{in-situ} capacitance measurements upon immersion in bulk binary mixtures 5CB$_{\rm x}$CH$_{\rm 1-x}$ with 5CB volume fraction $x$, as indicated in the figure. The gaps between the shaded time intervals indicate the times, when the sample cell is empty, see Fig.~\ref{fig:experimental_setup}(b). The capacitance $C_{\rm e}$ is measured immediately after removing of the bulk solution, see open circles with distinct colours. After this, the cell is again filled with the bulk binary mixture of 5CB$_{\rm x}$CH$_{\rm 1-x}$ with an increased volume fraction $x$.}
    \label{fig:adsorption_5CB_binary_mixture}
\end{figure}

In Fig.~\ref{fig:experimental_setup} a schematics of the experimental cell is shown. It consists of a syringe body. The mesoporous sample is fixed between two spring electrodes. A serious problem in such experiments can be stray (parasitic) capacitances. For this reason our measurements have been done with and without the binary solution, see panel (a) and (b) of Fig.~\ref{fig:experimental_setup}, respectively. The adsorption process takes place in the liquid-filled cell and can be monitored \textit{in-situ} by time-dependent capacitance measurements of the membrane $C(t)$, see Fig.~\ref{fig:adsorption_5CB_binary_mixture}. This allows one to monitor for each distinct bulk concentration $x$ the adsorption kinetics and in particular the reaching of an adsorption equilibrium in pore space. 

\begin{figure}
    \centering
    \includegraphics[width=0.95\textwidth]{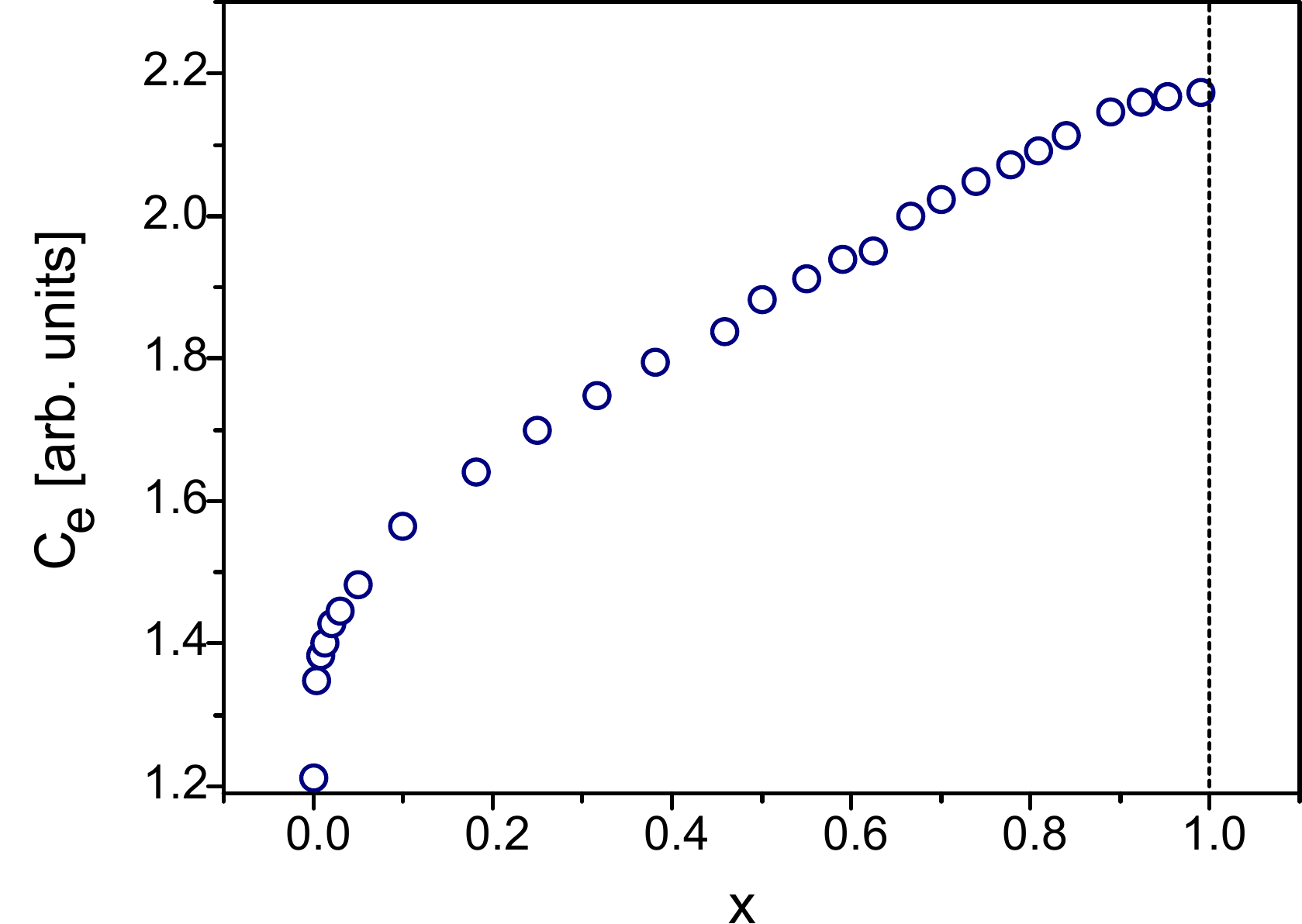}
    \caption{Measured equilibrium capacitance, $C_{\rm e}$ vs 5CB concentration $x$ for a mesoporous membrane pSiO$_2$ immersed into a binary mixture 5CB$_{\rm x}$CH$_{\rm 1-x}$.}
    \label{fig:equilibrium_capacitance}
\end{figure}

After equilibration the liquid cell is emptied and the capacitance $C_{\rm e}$ is measured, see open symbols in Fig.~\ref{fig:equilibrium_capacitance}. Thereby we have the same stray capacitance contribution in each measuring point. The resulting equilibrium capacitance, $C_{\rm e}$ vs $x$ for the nanoporous membrane pSiO$_2$ immersed into binary mixture 5CB$_{\rm x}$CH$_{\rm 1-x}$ is shown in Fig.~\ref{fig:equilibrium_capacitance}. After measuring $C_{\rm e}$ a new bulk solution 5CB$_{\rm x}$CH$_{\rm 1-x}$ was inserted and the procedure repeated as a function of increasing $x$. 

To calculate the adsorption isotherm from the capacitance isotherm certain assumptions have to be made. We anticipate that $C_{\rm e}$ in the equilibrium state scales linearly with the fractional volume content $V^*(x)=V(x)/V_0$ of 5CB in the pore volume, $V_0$. The fractional volume content consists of the molecules located both in the adsorbed 5CB layer at the pore walls plus molecules 5CB located in the binary mixture of the core region of the pore filling. The second assumption is that the volume concentration of the 5CB molecules inside the core region of the pore filling is the same as in the bulk solution surrounding the porous matrix, i.e.\ it is equal to $x$. Accordingly, the following equations are valid:
\begin{equation}
  V^*(x)=\frac{C_{\rm e} (x)-C_{\rm e} (x=0)}{C_{\rm e} (x=1)-C_{\rm e} (x=0)} \quad ,  
\end{equation}
\begin{equation}
    V(x)=f\,V_{\rm 0}+(1-f)\, V_{\rm 0}\, x \quad ,
\end{equation}

\begin{figure}
    \centering
    \includegraphics[width=0.95\textwidth]{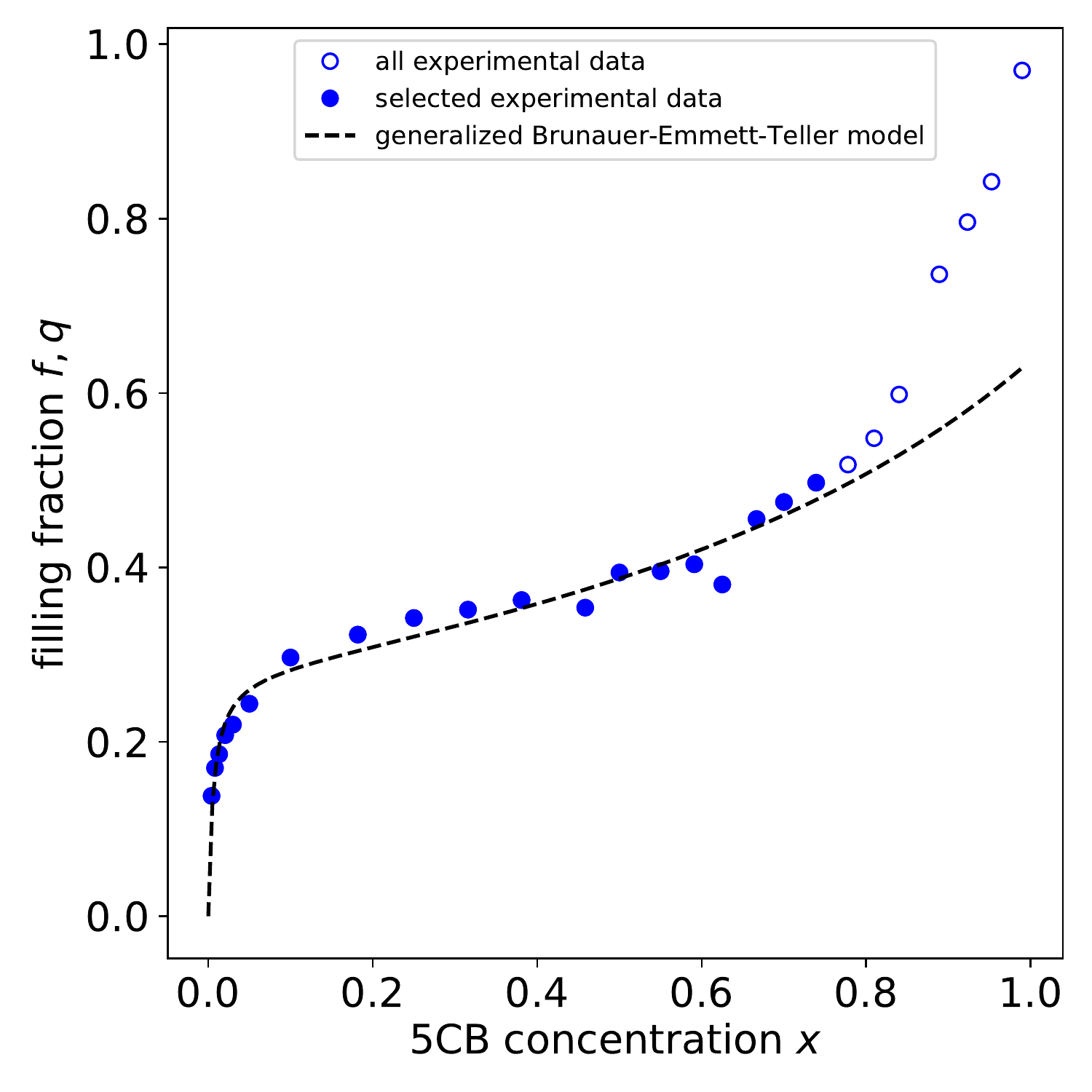}
    \caption{Adsorption isotherm of 5CB in mesoporous silica. Plotted is the fractional 5CB filling adsorbed from binary mixtures 5CB$_{x}$CH$_{1-x}$ as inferred from electrical capacitance measurements (blue markers). The dashed black line corresponds to a fit using the generalized BET model (Eq. \ref{eq:generalized_BET-equation}). The solid markers were used to fit the experimental data.}
    \label{fig:experimental_adsorption_isotherm_5CB}
\end{figure}

Thus for the fractional filling of adsorbed 5CB as a function of $x$, $f(x)$ follows:
\begin{equation}
    f(x)= \frac{V^*(x)-x}{1-x}
    \label{eq:filling_fraction}
\end{equation}
The filling fraction $f$ versus concentration $x$ is shown in Fig.~\ref{fig:experimental_adsorption_isotherm_5CB}. An unknown parameter in such a calculation is the $C_e(x=1)$ value [see eq.~\ref{eq:filling_fraction}], since at $x=1$ the anisotropic nematic LC phase is expected inside the pores at T=296\,K. Accordingly, our last data point corresponds indeed to x=0.99. The hypothetic value $C_e(x=1)$, which is quite close to $C_e(x=0.99)$ has been obtained (precisely adjusted) assuming that $f(x) \rightarrow 1$ at $x \rightarrow 1$.

 The 5CB adsorption isotherm in Fig.~\ref{fig:experimental_adsorption_isotherm_5CB} is similar to the type II gas adsorption isotherm \cite{Thommes2015}, and thus at low coverage can be described by the Brunauer-Emmett-Teller (BET) equation. For adsorption of a solute in a liquid medium the BET equation was generalized as follows \cite{Gritti2003}:
\begin{equation}
    q^*=q_{\rm m} \frac{b_{\rm s} x}{(1-b_{\rm l} x)(1-b_{\rm l} x+b_{\rm s} x)}
    \label{eq:generalized_BET-equation}
\end{equation}
where $q^*$ is the adsorbed amount, $q_{\rm m}$ is the monolayer capacity, $x$ is the solute concentration, $b_{\rm s}$ and $b_{\rm l}$ are the equilibrium constants for solute-adsorbent and solute-solute interactions respectively. This equation has been widely used for describing adsorption of various solutes \cite{Ebadi2009}, e.g.\ in a recent work of the Findenegg group it was applied for modelling adsorption of proteins onto silica nanoparticles in aqueous media \cite{Meissner2015}.

We applied Eq.~\ref{eq:generalized_BET-equation} to fit the 5CB experimental adsorption isotherm. Note that we expressed the adsorbed amount $q$ here in units of the complete filling, i.e. $q=f=1$ corresponds to a complete filling of the pores. Unlike the original BET equation, Eq.~\ref{eq:generalized_BET-equation} is a three-parametric dependence, thus we used the Trust Region Reflective algorithm, as implemented in SciPy \cite{Virtanen2020} constraining the parameters as non-negative and using various initial guesses. We also varied the interval of concentrations used for the fitting, and found that the solution for the monolayer capacity $q_{\rm m}$ changes insignificantly for the range of concentrations between 0.35 and 0.8. The solution for the interval [0, 0.75] gives $b_{\rm l} = 0.560$; $b_{\rm s} = 169.5$; $q_{\rm m} = 0.281$. This solution is plotted as a dashed line in Fig.~\ref{fig:experimental_adsorption_isotherm_5CB} and shows a nice fit within the considered interval.  

Beyond this interval the fit underestimates the amount adsorbed which suggest that a pore-filling beyond film-growth at the pore walls takes place, similar to the capillary condensation in the gas adsorption. Finally, assuming the pore geometry as cylindrical we recalculated the filling fraction for the monolayer $f_{\rm m} = q_{\rm m}$ to the thickness of the monolayer $h_{\rm m}$ using
\begin{equation}
    f_{\rm m} = \frac{R^2 - \left( R - h_{\rm m} \right)^2 }{R^2},
    \label{eq:monolayer}
\end{equation} 
where $R = D/2$ is the pore radius. Taking $R = {5.4}\,{nm}$, Eq.~\ref{eq:monolayer} gives the monolayer thickness $h_{\rm m}= {0.8}\,{nm}$. The 5CB molecule has a length of 1.9~nm and a diameter of 0.7~nm \cite{Leadbetter1975, Lay2000}. Thus, presumably the 5CB molecules are lying flat on the silica surfaces, similarly as it has been inferred for 8CB on planar quartz surfaces by Brewster-angle reflection ellipsometry and surface optical second harmonic generation \cite{Drevensek2003}. 

These conclusions with regard to the adsorption behaviour are consistent with optical measurements \cite{Huber2013} as well as dielectric spectroscopy studies upon sequential filling of mesoporous silica with 7CB from binary 7CB/acetone solutions  \cite{Sinha1997, Sinha1998,Calus2015a, Calus2015b}. In those experiments also first a regime with monolayer formation at the pore walls and a subsequent formation of capillary bridges was inferred from distinct optical signatures (changes in the optical birefringence and light scattering) as well as distinct molecular mobilities of these two molecular population in pore space. In particular, it was found that the first adsorbed layers at the pore walls are significantly slower in their molecular reorientation dynamics. The study presented here supports these functional differences by a rather thermodynamic distinction between the two populations.   

\section{Conclusions}
We presented an experimental study on the adsorption of 5CB from binary 5CB/methanol solutions in mesoporous silica by capacitance measurements. The resulting adsorption isotherm is reminiscent of classical gas isotherms of type II indicating multilayer growth on pore walls with heterogeneous LC-wall interactions and can be described by a generalized BET-isotherm, adapted for adsorption from binary solutions. Deviations from the BET description at higher filling fractions, higher concentrations, respectively, indicate a transition from LC-film growth at the pore wall to filling of the pore centre reminiscent of the capillary condensation transition of gases in mesoporous media. The monolayer capacity derived from the sorption-isotherm analysis hints towards the formation of a layer with homogeneous LC anchoring at the pore walls, i.e. the first 5CB monolayer is lying flat on the silica surface.

We hope that our study will stimulate complementary experimental and theoretical studies on adsorption of LC molecules from binary solutions. \textcolor{black}{The technique presented along with the simple analysis schemes can be employed also for other molecular systems with low vapour pressures, like discotic liquid crystals. Its applicability is also independent of the mesogenic (liquid-crystalline) properties, necessitates however a good liquid solvent. Thus the experimental approach is very versatile and is suitable for many molecular systems.} From a more general perspective our techniques allows one to prepare LC-mesoporous hybrid materials which could be of particular interest for photonic \cite{Spengler2018, Huber2020} and organic electronic applications \cite{Duran2012, Huber2020}. 

\section{Acknowledgment}
GG and PH dedicate this work to Professor Gerhard  Findenegg (Technical University Berlin), a pioneer in the field of molecular adsorption in porous materials and self-assembly in confinement. Funding by the Deutsche Forschungsgemeinschaft (DFG, German Research Foundation) Projektnummer 192346071, SFB 986 “Tailor-Made Multi-Scale Materials Systems” and the DFG Graduate School GRK 2462 ”Processes in natural and technical Particle-Fluid-Systems (PintPFS)” (Projektnummer 390794421) is gratefully acknowledged. The presented results are part of a project that has received funding from the European Union’s Horizon 2020 research and innovation programme under the Marie Skłodowska-Curie grant agreement No. 778156. Support from resources for science in the years 2018-2022 granted for the realization of  international co-financed project Nr. W13/H2020/2018 (Dec. MNiSW 3871/H2020/2018/2) is also acknowledged.

%\bibliographystyle{unsrt}
%\bibliography{bibliography}

\end{document}